\documentclass{an}
\usepackage{graphicx}
\usepackage{times}
\usepackage{fancyhdr}
\begin{document}

\title{Unbiased Studies of Diffuse Extragalactic Radio Sources}

\author{Lawrence Rudnick, Kisha M. Delain \and Jeffrey A.  Lemmerman}
\institute{University of Minnesota, 116 Church St SE, Minneapolis, MN  55455}

\date{Received $<$date$>$;
accepted $<$date$>$;
published online $<$date$>$}

\abstract{We present an overview of unbiased studies of diffuse extragalactic radio sources.
We use a previously developed filtering technique to remove compact sources from large
surveys such as WENSS, WISH, and NVSS and examine the residual diffuse emission.  A search
of these residuals, unbiased by optical or X-ray selection, has uncovered a wide variety of
diffuse sources, including those associated with groups of galaxies, ``blank'' fields, and
previously unrecognized diffuse radio galaxy emission, as well as halos and relics of rich
clusters.  A second, targeted survey of the brightest X-ray clusters results in a number of
new sources, and illuminates some important selection effects.   When the well-established
relationship between X-ray and radio halo luminosities is extrapolated to lower levels,
the apparent correlation holds quite well for any type of diffuse source, even
background noise.  The observed correlations must therefore be scrutinized for possible
selection effects before physical interpretations are made.  We briefly mention  a
wide-field mapping project to search for large-scale structures, and goals for the next
generation of studies in magnetic field - angular size parameter space.}


\correspondence{larry@astro.umn.edu}

\maketitle

\section{Introduction}
A great deal of effort has gone into the study of diffuse radio sources in clusters of galaxies,
 i.e., sources not obviously associated with a parent AGN.  The growing number of these sources
 has led to the identification of different classes, including centrally located radio halos
 and mini-halos, and radio ``relics'' of various types (e.g. Kempner et al. 2004).  Although
 cluster searches have uncovered this rich variety of phenomena, there have not been surveys
 done in a quantitative way, with well-defined upper limits for non-detections.  In addition,
 unbiased surveys for diffuse emission, independent of the optical or X-ray environment, have
 not been systematically carried out.  We have therefore begun to address these issues with a
 series of studies which we preview here.  A companion paper (Delain and Rudnick, 2005) shows
 some of our early results on diffuse sources in groups and poor clusters of galaxies.

\section{WENSS/WISH searches}
We applied the multi-resolution filtering technique developed by Rudnick (2002) to the
WENSS and WISH surveys (Rengelink et al. 1997, De~Breuck et al. 2002)
to remove the contributions from compact
sources.  At this point, we have simply examined the residual large scale emission by eye,
and identified strong diffuse structures.  In this examination, we find a rich variety of
sources, including extended radio galaxies, previously known radio halos and relics, the
occasional supernova remnant or HII region when we venture too close to the galactic plane, and
new diffuse sources that we are in the process of identifying.

\begin{figure*}
\resizebox{\hsize}{!}
{\includegraphics[]{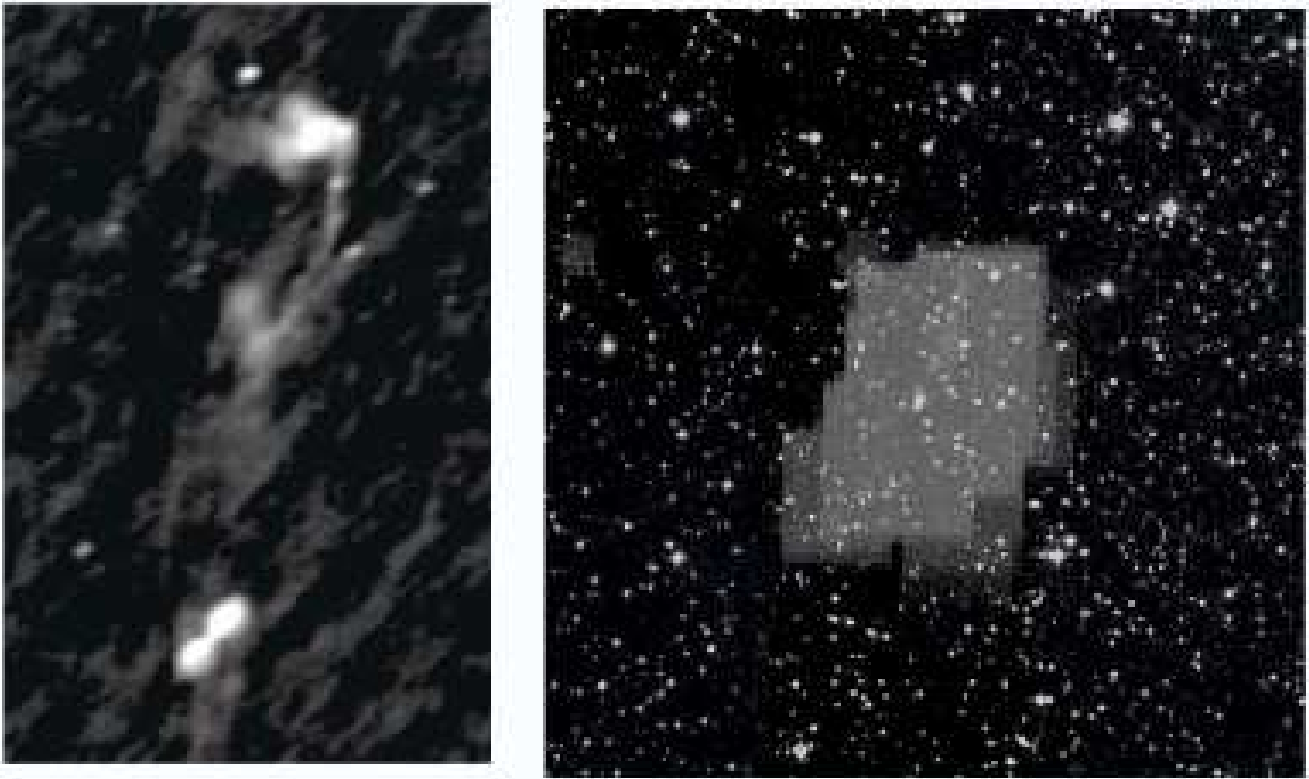}}
\caption{Left:  Deep VLA image at 1.4 GHz of WAT source discovered in blind search of WISH. Field height
is $\approx$ 45 arcmin.  Right: Filtered 327 MHz  WENSS image showing diffuse emission in a ``blank''
field, about 30 arcmin across. Rectangular patches are the result of filtering.}
\label{wish}
\end{figure*}

 Two very different examples of newly detected sources are shown in Figure 1.  The source on the left
 was seen in the filtered WISH images as a group of diffuse patches of uncertain geometry.  We
 conducted follow-up observations on NRAO's Very Large Array \footnote {The National Radio Astronomy
 Observatory is a facility of the National Science Foundation operated under cooperative agreement
 by Associated Universities, Inc.}, which are shown here.  The most prominent source in the upper
 part of the image is a very large wide-angle tail (WAT).  It extends for 500 kpc between the two sharp bends
 to the east into the diffuse lobes.  The jets of a second possible WAT are seen in superposition.  This
 region is located on the outskirts of the Ophiucus cluster of galaxies, near the Sagittarius group.
 Thus, this source is one of the few examples where the medium between clusters can be probed.

 The source on the right of Figure 1 has an extent of approximately 20\arcmin, and is not apparently
 associated with any optical galaxy, group, or cluster.  No ROSAT emission is detected from this
 location.  No matter what the eventual identification, this source is likely to be unique -- whether
 it's a physically small detached piece of an old radio galaxy or the extremely large and luminous
 halo of a distant unseen cluster.

 \section{Quantitative search of brightest X-ray clusters}

We conducted an objective survey of the 88 clusters from the ROSAT Brightest Cluster Sample (Ebeling
et al. 1996) that are in the declination range of the WENSS survey.  Images of the region around each
cluster were first filtered to remove compact sources.  Then, the remaining flux within a circle of
radius 500 kpc was determined after correction for a local background.  A control field was set up
two degrees away in declination for each cluster, in order to determine the background noise level.
The same respective redshifts were assumed for the control fields.  The upper envelope of control
field fluxes as a function of redshift was used to establish a 95\% confidence level.  Twenty-seven
detections
of diffuse emission were made, most of which were known previously, and included both extended
radio galaxies as well as cluster halos.  Several new diffuse sources were detected, including one
in Abell 2061, shown in Figure 2.  The bright patch to the SW of the cluster is the relic identified
by Kempner \& Sarazin (2001). The diffuse emission superposed on the X-ray emission from Abell 2061
has not been noticed previously, and peaks around the location of the shock identified with
 {\it BeppoSAX} (Marini et al. 2004).

\begin{figure}
\resizebox{\hsize}{!}
{\includegraphics[]{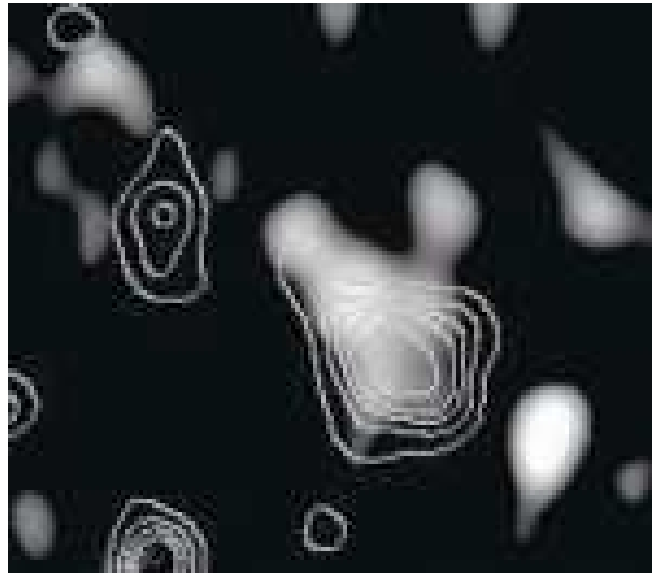}}
\caption{Diffuse radio emission in the region of the A2061 cluster from the filtered WENSS images shown
as greyscale, overlaid by
contours of the broadband ROSAT emission.  Both images have been convolved to 300 arcsec.}
\label{a2061}
\end{figure}

Figure 3 shows a plot of radio vs. X-ray luminosity for several different samples.  The rich
cluster halos summarized by Bacchi et al. 2003) are the highest luminosities on average, and our
new 99\% confidence detections are
shown as filled circles.  These new detections include a variety of diffuse sources, including extended radio
galaxies, which also fall on the same luminosity trend.  The open circles, extending the
same correlation to luminosities several orders of magnitude below rich clusters, are from the
{\it control} fields, and are therefore noise.  (We used the absolute value of the flux within
the 500 kpc radius circle for the control fields.)

Inferences about radio/X-ray scaling relationships must therefore be made with caution.   Bacchi et al. 2003 claim a  strong correlation between radio monochromatic power at 1.4 GHz and bolometric X-ray luminosity only for merging clusters with $>$ 1 Mpc radio halos.  However, using a radio aperture of 1 Mpc,  we find the same correlation  for the brightest X-ray clusters, selected without regard to their merger status or whether the radio emission comes from halos or diffuse radio galaxies. A second problem comes from the absence of published upper limits. Feretti (2004) notes that it is difficult to include upper limits because the size of the (undetected) source is unknown, and she suggests using radio surface brightnesses instead.  Here  we are able to illustrate the sensitivity selection effects by including `pseudo-measurements' on blank fields offset from actual X-ray clusters, using fixed 1 Mpc apertures at the X-ray cluster redshift.  The result is an apparent (spurious) extension of the correlation below $10^{23.5}$ W/Hz.  We therefore agree with Feretti (2004) that even low X-ray luminosity clusters may host low radio luminosity halos, but that the next generation of sensitive radio instruments, such as LOFAR and SKA, will be needed to decide this important issue.   

\begin{figure}
\resizebox{\hsize}{!}
{\includegraphics[]{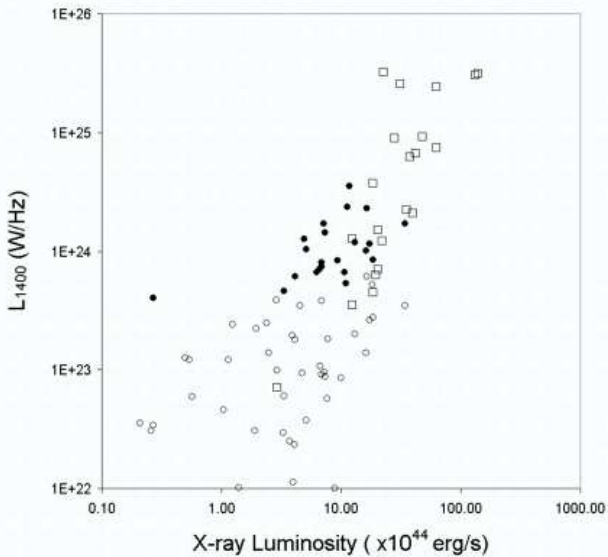}}
\caption{Radio vs. X-ray luminosities for several samples, as described in more detail in the text.
Open squares: cluster halos. Filled circles: diffuse cluster sources from current survey.  Open
circles: control fields.}
\label{correlations}
\end{figure}

\section{Searching for large scale magnetic fields}
We have been mapping a 36 square degree region in the TONS-08 field (Brand et al. 2003, 2005)
at 330 MHz
using the VLA in its D and B configurations, in order to search for filamentary or other
large scale emission.  We cover the field with 22 different pointings.
At present, our sensitivity is limited by the accuracy of removing
emission from compact sources, especially those seen several degrees off axis.  This removal
is complicated not only by ionospheric gradients but the asymmetries in the primary beams of
the telescope which changes the response to off axis sources as a function of parallactic angle.

The preliminary analysis shows the challenges inherent in
looking for diffuse emission.  The left side of Figure 4 shows a filamentary network of
diffuse features (with white boundaries drawn to guide the eye), overlaid by the compact
sources that have been removed using our filtering algorithm.
 The filamentary features still trace the distribution of the (subtracted) compact sources.
 This is likely due to two different causes:  (1) small amounts of diffuse emission around the
 compact sources that survive the filtering procedure, and (2) a large number of faint
 compact sources that aggregate in the same regions (as expected from the two-point
 correlation function for radio sources, e.g. Overzier et al. 2003) and are not well removed.
 The right side of Figure 4 shows the
 result of a stronger filtering of compact emission (filtering out close blends and very
 slightly extended emission), and the associated filaments are now gone.  The remaining
 diffuse structures are still consistent with noise.  These results highlight the
 importance and difficulty of removing contributions from compact sources in searches for
 extended emission.

\begin{figure}
\resizebox{\hsize}{!}
{\includegraphics[]{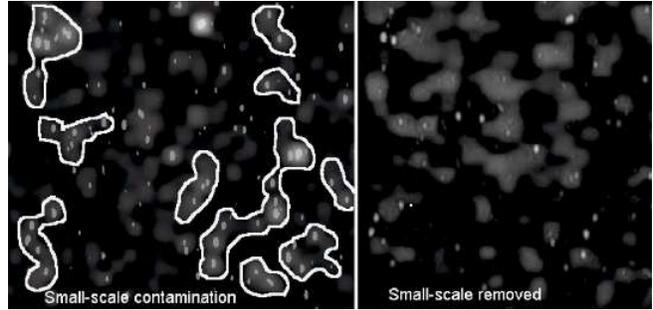}}
\caption{Compact and (pseudo-) filamentary emission in at 330 MHz from a VLA wide-field
mapping project.  The images shown are $\approx$ 1 degree across.}
\label{filaments}
\end{figure}

Figure 5 summarizes, in the phase space of magnetic field strength vs. angular size, the
current state of our observations of diffuse magnetic fields.  A horizontal line shows
the value of a magnetic field that would produce a pressure equivalent to the
 thermal pressure in rich clusters of galaxies.  The minimum energy fields of cluster
 radio galaxies are comparable to this number, although their diffuse regions, and
 those of cluster halos often fall below the thermal pressure value.  The shaded region
 around a few $\mu$~G (labeled $\tau_{max}$) shows an important selection effect. Sources
 with magnetic fields much stronger than this would live for shorter times because of
 increased synchrotron losses, as long as reacceleration of relativistic particles is not
 important.  Sources with weaker fields would also live for a shorter
 time (at fixed observing frequency).  This is because we would be observing higher energy electrons
 at the fixed observing frequency, and these higher energy electrons would lose energy quickly through
 inverse Compton emission off the cosmic background.

Weak magnetic fields may be quite common, and important to the evolution of structure
in the universe.  In terms of detectability, the important region to focus on is the
one indicated by the spiky ellipse, where the studies described here are beginning to probe.
A small circle, at slightly smaller sizes, indicates the diffuse source in Abell 2061,
as shown in Figure 2.  At the bottom of the figure we see the WHIM, plotted at a pressure
matched field strength.  This is the next frontier for magnetic field studies,
around 0.1$\mu$~G, if we are to be sensitive to structures outside of rich clusters.
The curved line extending through the WHIM region is the limit set by confusion -- on
small scales by background sources and on large scales by the Milky Way.  Techniques
to push below these confusion limits are being developed, and will eventually allow a
much fuller study of the rich structure of the cosmic web.
\begin{figure*}
\resizebox{\hsize}{!}
{\includegraphics[]{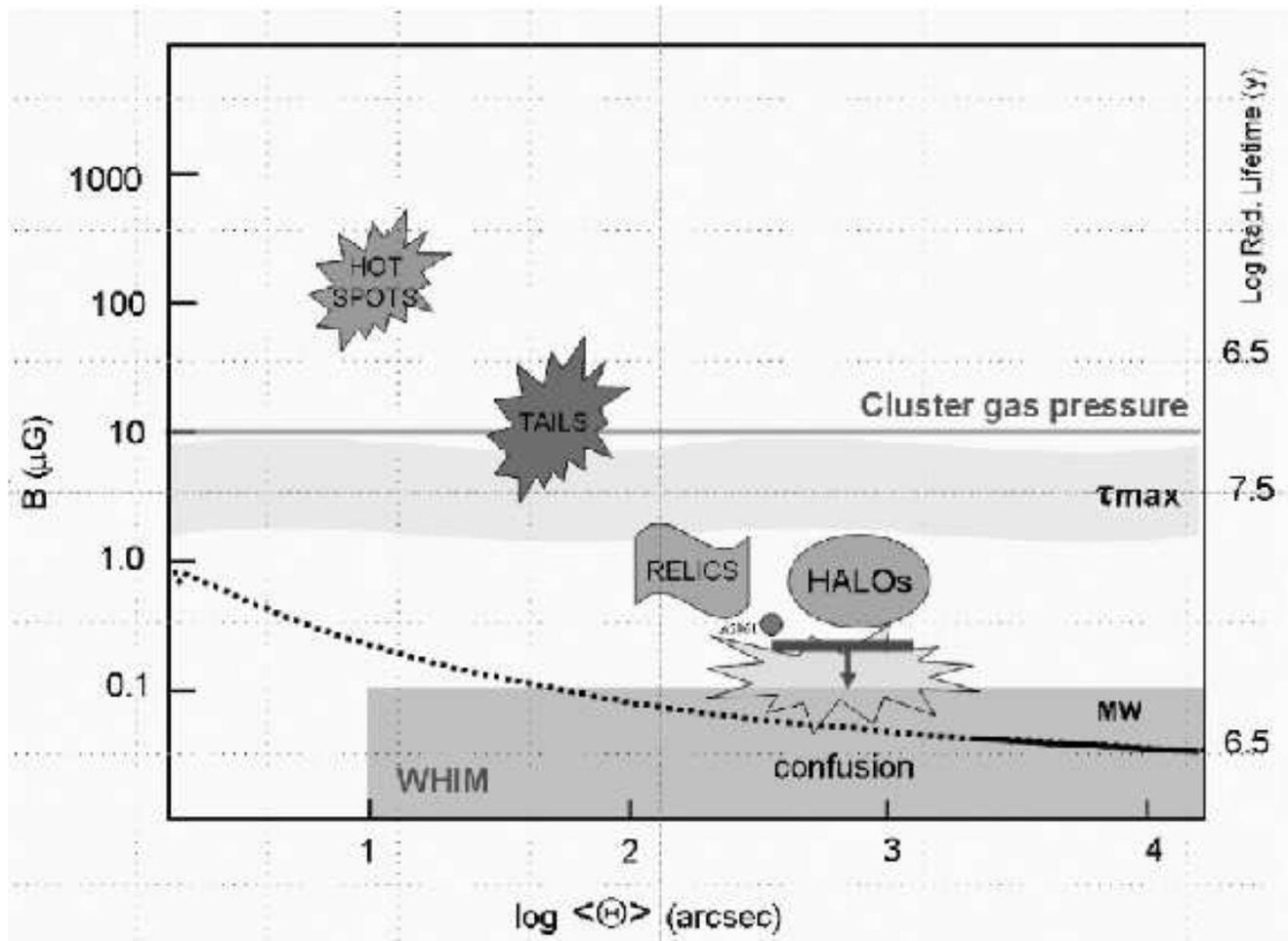}}
\caption{Magnetic field strength, Angular Size phase space of diffuse magnetic fields, showing
known objects, target sensitivities and observational constraints.}
\label{btheta}
\end{figure*}

\acknowledgements
We acknowledge the use of the WENSS and WISH archives, and thank Carlos de Brueck
for making the WISH data available. We have made use of the ROSAT Data Archive of the 
Max-Planck-Institut für extraterrestrische Physik (MPE) at Garching, Germany.
Partial support for this research comes from the US National Science Foundation 
grant 03-07600 to the Univ. of Minnesota.\\

\end{document}